\documentclass[sigconf]{acmart}

\acmYear{2019}\copyrightyear{2019}
\setcopyright{rightsretained}
\acmConference[IMC '19]{IMC '19: ACM Internet Measurement Conference}{October 21--23, 2019}{Amsterdam, Netherlands}
\acmBooktitle{IMC '19: ACM Internet Measurement Conference, October 21--23, 2019, Amsterdam, Netherlands}
\acmPrice{}
\acmDOI{10.1145/3355369.3355575}
\acmISBN{978-1-4503-6948-0/19/10}

\usepackage{graphicx}
\usepackage{booktabs}
\usepackage{hyperref}
\usepackage{threeparttable}

\usepackage{pifont}
\newcommand{\cmark}{\ding{51}}%
\newcommand{\xmark}{\ding{55}}%

\begin{document}

\title{An Empirical Study of the Cost of DNS-over-HTTPS}

\author{Timm~Boettger, Felix~Cuadrado, Gianni~Antichi, Eder~Leao~Fernandes,\break Gareth~Tyson, Ignacio~Castro and Steve~Uhlig}
\affiliation{
	\institution{Queen Mary University of London}
}
\renewcommand{\shortauthors}{Timm Böttger et al.}

%TC:break Abstract
\begin{abstract}
DNS is a vital component for almost every networked application.
Originally it was designed as an unencrypted protocol, making user security a concern. 
%Originally however it was designed as an unencrypted protocol yielding concerns about user security. 
%concerns about user security however have driven researchers and practitioners to look for more secure approaches.
%Indeed, by just examining a user's DNS requests one can already build very specific user profiles. %infer the applications being used or the websites being visited.
%This for example allows to build very specific user profiles.
%Although, the idea of running DNS-over-TLS (DoT) has already been around for a few years now, as of today, DoT is not widely supported by DNS resolvers.
%Instead, the more recently proposed DNS-over-HTTPS (DoH) protocol has received much more attention. 
%Specifically, The idea and standardization of a secure DNS protocol (DNS-over-TLS (DoT)) have already been around for a few years now.
%However, even as of today, DoT is not widely supported by DNS resolvers. Instead, the more recently proposed DNS-over-HTTPS 
%(DoH) protocol has received much more attention. Among others, Google, Cloudflare and Mozilla are supporting and actively pushing for it.
DNS-over-HTTPS (DoH) is the latest proposal to make name resolution more secure.

In this paper we study the current DNS-over-HTTPS ecosystem, especially the cost of the additional security.
We start by surveying the current DoH landscape by assessing standard compliance and supported features of public DoH servers.
We then compare different transports for secure DNS, to highlight the improvements DoH makes over its predecessor, DNS-over-TLS (DoT).
These improvements explain in part the significantly larger take-up of DoH in comparison to DoT.
%In this paper we first compare DoT and DoH to shed some lights on why the latter has attracted more interest by Google, Cloudflare and Mozilla, among others.
%We then conduct a multiple measurements to stud yand characterize the DoH landscape.

Finally, we quantify the overhead incurred by the additional layers of the DoH transport and their impact on web page load times.
We find that these overheads only have limited impact on page load times, suggesting that it is possible to obtain the improved 
security of DoH with only marginal performance impact.

%We conclude this study by quantifying the overheads the DoH transport causes
%We look at standard compliance, supported features of public DoH servers and the associated overheads when moving from UDP transport to HTTPS.
%Finally, we assess its impact on web page loading times.
%We also quantify the overheads caused by the switch from a UDP transport to HTTPS.
%Further, we assess the impact the switch to DoH has on web page loading times.

%We conclude the paper by discussing the expected implications of DoH on the DNS landscape.

%We conclude by comparing resolution latency of these early DoH servers against 
%regular DNS servers to get a first feeling for the response times DoH might 
%be able to achieve.
%\keywords{DNS \and Transport}
\end{abstract}
%TC:break _main_

%%
%% The code below is generated by the tool at http://dl.acm.org/ccs.cfm.
%% Please copy and paste the code instead of the example below.
%%
%%
%% Keywords. The author(s) should pick words that accurately describe
%% the work being presented. Separate the keywords with commas.
% \tkeywords{DNS, transport}

\begin{CCSXML}
<ccs2012>
<concept>
<concept_id>10003033.10003039.10003048</concept_id>
<concept_desc>Networks~Transport protocols</concept_desc>
<concept_significance>500</concept_significance>
</concept>
<concept>
<concept_id>10003033.10003079.10011704</concept_id>
<concept_desc>Networks~Network measurement</concept_desc>
<concept_significance>500</concept_significance>
</concept>
<concept>
<concept_id>10003033.10003083.10003014</concept_id>
<concept_desc>Networks~Network security</concept_desc>
<concept_significance>500</concept_significance>
</concept>
</ccs2012>
\end{CCSXML}

\ccsdesc[500]{Networks~Transport protocols}
\ccsdesc[500]{Networks~Network measurement}
\ccsdesc[500]{Networks~Network security}

\keywords{DNS-over-HTTPS, Transport, Performance}

\maketitle
\section{Introduction}

\begin{figure}
	\centering
	\includegraphics[width=\columnwidth]{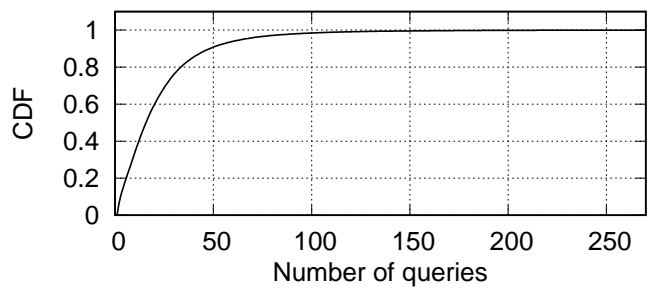}
	\caption{CDF of the number of DNS queries required to retrieve all embedded objects for each of the top 100k Alexa sites.}
	\label{fig:cdf-queries-pages}
\end{figure}

Introduced in 1983, the Domain Name System (DNS) has become a critical component of the Internet.
%DNS is a hierarchical decentralized naming system, with the primary purpose of resolving domain names to IP addresses.
%DNS queries have typically consisted of a single UDP request from the client and a single reply from the server.
%
%DNS plays a vital role for many networked applications today.
In addition to its original purpose of domain name resolution, DNS has also gained relevance due
to its intensive use by Content Distribution Networks (CDNs) for traffic redirection~\cite{bottger2018open,calder2013mapping}.
Most websites nowadays include content from third parties, hence requiring multiple DNS queries~\cite{DBLP:conf/imc/ButkiewiczMS11} to access a single page. To highlight this, Figure~\ref{fig:cdf-queries-pages} shows the number of DNS queries required to fully retrieve each page in the Alexa global top 100k sites. Each website was retrieved through Firefox, logging the DNS requests at the stub resolver. Caches of both Firefox and the DNS stub resolver were emptied before requesting the next website.
The figure illustrates that multiple DNS queries per page are the norm rather than the exception: about 50\% of the sites require at least 20 DNS queries.
%
%DNS has become a key performance and security factor for networked applications. Previous studies have shown that DNS is an important factor for the latency of applications~\cite{bozkurt2017internet}, with DNS resolution times varying drastically depending on the location and DNS provider.

DNS impacts networked application performance~\cite{bozkurt2017internet} and can
reveal information about the destination of a connection~\cite{bortzmeyer2015dns}. Addressing increasing concerns about security, DNS-over-TLS (DoT)~\cite{rfc7858} and more recently DNS-over-HTTPS (DoH)~\cite{rfc8484} have been proposed within the IETF. To increase security, these protocols rely on a TLS session between the client and the resolver. In the case of DoH, this TLS session also contains a HTTP connection.
So far, DoT has only gained limited traction, whereas DoH has gathered substantial momentum already~\cite{apnicdoh}, with the help of notable players like Mozilla, Cloudflare and Google.

In this paper, we take a first look at the implications of securing DNS with DoH.
%, and in particular, considering current trends, at doing so using DoH.
We also compare DoT and DoH to shed some light on why the latter has recently gained so much interest.
%The potential advantages of DoH include i) leveraging the rich HTTPS landscape, e.g., existing libraries and tool sets, ii) the fact that it can be mixed with other HTTPS traffic, making its identification more difficult, and iii) that the integration of DoH in the web-browser might speed up adoption.
%
%In this paper we take a first look at the implications of securing DNS, and in particular at doing so using DoH.
The following are the main
contributions:
%For instance, since Dot and DoH provide similar security assurances, it is not clear yet what allowed DoH to gain traction significantly faster than DoT did.
%For instance, the discovery of a DoH server is still not defined\footnote{\url{https://datatracker.ietf.org/doc/draft-nottingham-doh-digests}}.
%However, the adoption of DoH by important players like Mozilla, Google, and Cloudflare means that there is critical mass behind DoH, so that DoH will likely become a relevant protocol in the wild. In this paper, we take a first look at this evolving DNS landscape.
%in the wild so that DoH will have critical implications for both security and performance. This paper is a first attempt to understand those implications.

%In this paper we take a first look at the performance implications of securing DNS, and in particular at doing so using DoH.
%Preliminary evaluations of DoH by Mozilla have shown a moderate negative impact in performance with --even large-- gains
%for those DNS transactions performing worst in traditional DNS\footnote{\url{https://blog.nightly.mozilla.org/2018/08/28/firefox-nightly-secure-dns-experimental-results/}}.
%Our findings show that....
%In this paper makes the following contributions:
\begin{enumerate}
	\item We survey and characterize the current landscape for secure DNS via HTTP and TLS.
	\item We compare different transport protocols for securing DNS resolution, to understand the momentum behind DoH.
	\item We quantify the overheads incurred by the additional HTTP and TLS layers in DoH.
	\item We take a first look at the impact that switching to DoH has on web performance, more specifically DNS resolution times and page load times.
\end{enumerate}

\section{The DoH landscape}

%DNS-over-HTTPS (DoH) is a relatively young protocol.
%From the major browsers, Mozilla Firefox is so far the only one with documented DoH support.
%Mozilla has also run a preliminary study on the effect of DoH on browsing experience~\cite{mozillablog}, demonstrating that its performance on their Cloudflare-based implementation is acceptable. Indeed, despite some general negative impact of the overhead on resolution time, the slowest resolution times can actually be better than with UDP. However, such results are too preliminary and limited to a specific implementation and infrastructure. Additional work needs to be done to genuinely understand the performance impact of DoH in the wild.

To better understand the current landscape of DoH resolvers, we take the list of DoH servers curated by the curl project,\footnote{\url{https://github.com/curl/curl/wiki/DNS-over-HTTPS}} and assess their supported feature set.
% Table~\ref{tab:doh-resolvers} lists the DoH resolvers and their URLs.
We initially retrieved all information in this section on 10 October 2018. We then verified it and, where necessary, updated entries in both tables again on 10 September 2019.

\begin{table}
\centering
\setlength\tabcolsep{2pt}
\small
\begin{tabular}{l|l|c}
\toprule
Provider & DoH URL & MK \\
\midrule
Google (i) & https://dns.google.com/resolve & G1 \\
Google (ii) & https://dns.google.com/dns-query & G2 \\
Cloudflare & https://cloudflare-dns.com/dns-query & CF \\
Quad9 & https://dns.quad9.net/dns-query & Q9 \\
CleanBrowsing & https://doh.cleanbrowsing.org/doh/family-filter & CB \\
PowerDNS & https://doh.powerdns.org/ & PD \\
Blahdns & https://doh-ch.blahdns.com/dns-query & BD \\
~ & https://doh-jp.blahdns.com/dns-query & ~ \\
~ & https://doh-de.blahdns.com/dns-query & ~ \\
SecureDNS & https://doh.securedns.eu/dns-query & SD \\
Rubyfish & https://dns.rubyfish.cn/dns-query & RF \\
Commons Host & https://commons.host/ & CH \\
\bottomrule
\end{tabular}
\vspace{\baselineskip}
\caption{Compared DoH resolvers. Markers (MK) refer to column identifiers used in Table~\ref{tab:comparison-of-features}. Blahdns offers DoH services on three different URLs.}
\label{tab:doh-resolvers}
\end{table}

As Table~\ref{tab:doh-resolvers} shows, major players like Google, Cloudflare and IBM (Quad9), as well as some smaller players, support DoH.
We observe diversity in their service configurations. While different base URLs for every service can be expected, it is surprising to see four different URL paths (\texttt{/}, \texttt{/resolve}, \texttt{/dns-query}, \texttt{/family-filter}) just among these nine providers.
% Steve: the path observation is shallow unless we explain how important it is. IMC reviewers won't find something like this interesting unless we make it so.
Google even uses different paths to two different services with the same base URL.
While technically the DoH RFC~\cite{rfc8484} does not mandate a specific path to be used and leaves it up to the service operators, the majority of services still use the path \texttt{/dns-query}, which is the one used in all examples in the RFC. Given the huge efforts spent by operators to obtain easy-to-remember and thus easy-to-configure IP addresses for their UDP based DNS servers~\cite{apniccloudflare,cloudflare1111}, seeing such a potentially confusing variety and choices for the DoH service parameters is noteworthy.
DoH operators indeed seem to have realised this, when we first collected this information in October 2018 we observed six different base paths for the same set of providers, while now we only observe four.

% Steve: not too convinced pushing this thing as interesting will fly.

We now examine the features supported by the individual resolvers. HTTP supports the transmission of different content types.
As per the DoH RFC, all DoH servers and clients must support the \texttt{application/dns-message} content type, which essentially is an encapsulation of the UDP DNS wireformat in HTTPS. Another widely supported type is \texttt{application/dns-json} which represents DNS messages in JSON format.
While a draft RFC for the JSON DNS format~\cite{bortzmeyer-dns-json-01} exists, its support is not mandatory for DoH servers.
The \texttt{application/dns-message} content type is supported by all implementations except Google's.
Google in fact operates two different services with two different paths (\texttt{/resolve} and \texttt{/dns-query}) on the same domain, with each service only supporting one content type.
Curiously enough, the service supporting the RFC mandated format was initially named \texttt{/experimental} and has since then being renamed to \texttt{/dns-query}.
This again highlights that operators have understood that too many different URLs are confusing and might lead to configuration errors.
Of the remaining eight providers four also support the JSON format on the same path as the DNS wireformat.
% Steve: For the curiously, do we have some explanation/guess?

\begin{table}
\setlength\tabcolsep{2pt}
\centering
\small
\begin{threeparttable}
\begin{tabular}{l|cccccccccc}
\toprule
Feature & G1 & G2 & CF & Q9 & CB & PD & BD & SD & RF & CH \\
\midrule
\texttt{dns-message} & \xmark & \cmark & \cmark & \cmark & \cmark & \cmark & \cmark  & \cmark & \cmark & \cmark\\
\texttt{dns-json} & \cmark & \xmark & \cmark & \cmark & \xmark & \xmark & \cmark  & \xmark & \cmark & \xmark \\
\midrule
TLS 1.0 & \xmark & \xmark & \cmark & \xmark & \xmark & \cmark & \xmark & \cmark & \cmark & \xmark\\
TLS 1.1 & \xmark & \xmark & \cmark & \xmark & \xmark & \cmark & \xmark & \cmark &\cmark & \xmark\\
TLS 1.2 & \cmark & \cmark & \cmark & \cmark & \cmark & \cmark & \cmark & \cmark & \cmark & \cmark\\
TLS 1.3 & \cmark & \cmark  & \cmark & \cmark & \xmark & \cmark & \cmark & \cmark & \xmark & \cmark\\
\midrule
CT & \cmark & \cmark & \cmark & \cmark & \cmark & \cmark & \cmark & \cmark & \cmark & \cmark \\
DNS CAA & \cmark & \cmark & \xmark & \xmark & \xmark & \xmark & \xmark & \xmark & \xmark & \xmark\\
OCSP MS & \xmark & \xmark & \xmark & \xmark & \xmark & \xmark & \xmark & \xmark & \xmark & \xmark \\
\midrule
QUIC & \cmark & \cmark & \xmark & \xmark & \xmark & \xmark & \xmark & \xmark & \xmark & \xmark \\
\midrule
DNS-over-TLS & \cmark & \cmark & \cmark & \cmark & \cmark & \xmark & \xmark & \xmark & \xmark & \xmark  \\
\midrule
Traf. Steering & DL\textsuperscript{*} & DL\textsuperscript{*} & AC\textsuperscript{+} & AC\textsuperscript{+} & AC\textsuperscript{+} & UC\textsuperscript{±} & UC\textsuperscript{±} & UC\textsuperscript{±} & UC\textsuperscript{±} & AC\textsuperscript{+} \\
\bottomrule
\end{tabular}
\begin{tablenotes}[para]\footnotesize
\item[*] DNS Load Balancing \item[+] Anycast \item[±] Unicast
\end{tablenotes}
\end{threeparttable}
\vspace{\baselineskip}
\caption{Comparison of DoH resolver features. Column names refer to markers in Table~\ref{tab:doh-resolvers}.}
\label{tab:comparison-of-features}
\end{table}

DoH was designed as a secure service with transport encryption via TLS. TLS support is thus a strict requirement.
There has been significant change on the TLS front recently, with TLS 1.3 becoming an official RFC and security vulnerabilities POODLE and BEAST rendering TLS 1.0 and lower standards insecure. %GlobalSign recommends to disable TLS 1.1 and only use TLS 1.2 and TLS 1.3~\cite{globalsigntls}.
All DoH servers support TLS 1.2, and seven of the nine providers also support TLS 1.3.
This is a positive sign towards broader acceptance of DoH, since when we assessed these features in October 2018 only Cloudflare and SecureDNS supported TLS 1.3.
%Surprisingly, only Cloudflare and SecureDNS support the future-proof TLS version 1.3, while players as significant as Google or IBM do not offer TLS 1.3.
On the other hand, we also see that some servers still support the deprecated TLS versions 1.0 and 1.1.
We suspect the reason is that some operators are concerned about compatibility issues with older client libraries and thus also support older TLS versions.
While the client can always insist on negotiating a connection with TLS 1.2 or higher, it would make sense that operators also use this opportunity to provide secure DNS and simultaneously put pressure on dropping TLS versions 1.1 or lower.

Four of the servers surveyed (Google, Cloudflare, IBM, and PowerDNS) also support DNS-over-TLS~\cite{rfc7858}, the previous RFC for encrypting DNS requests using TLS. Despite having a three-year headstart over DoH, DoT has failed to gain significant traction compared to the previous specification. We will further explore both protocols in the next section.

\begin{figure*}[t]
   \centering
   \includegraphics[width=\textwidth]{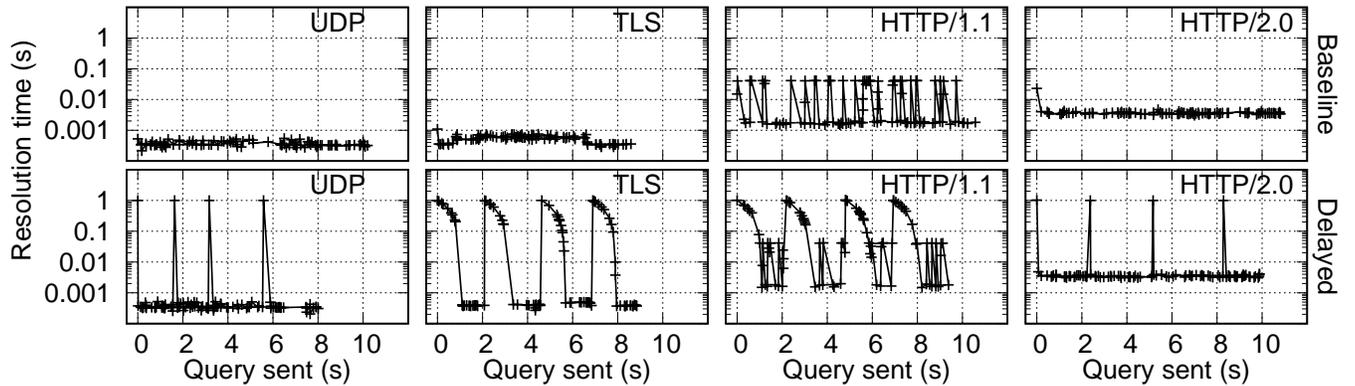}
   \caption{Impact of head-of-line-blocking on resolution times for DNS over different transport protocols. The upper charts depict the baseline and the lower ones the effect of a delay (1000ms for one in 25 queries).}
   \label{fig:hol-multiplot}
\end{figure*}

Finally, DNS-over-HTTPS relies on the PKI-certificate system to ascertain the identity of the DNS resolver.
To compensate known weaknesses and flaws of this system, techniques such as Certificate Transparency (CT)\footnote{\url{https://www.certificate-transparency.org}}, Certification Authority Authorization (CAA)~\cite{rfc6844} records and Online Certificate Status Protocol (OCSP)~\cite{rfc6960} have been proposed.
% Certificate transparency\footnote{\url{https://www.certificate-transparency.org}} is a Google-driven initiative to mitigate the effects of rogue certificate issuance.
% To this end, certificate authorities log all issued certificates to public CT logs.
% This allows the legitimate owner of a domain to check these logs and identify (rogue) certificates that were issued without their knowledge.
% The DNS certification authority record~\cite{rfc6844} allows domain owners to indicate which certificate authorities are legitimately allowed to issue certificates for domains. This counteracts the ability of any trusted CA to issue certificates for any domain.
% The Online Certificate Status Protocol~\cite{rfc6960} is used to actively check whether a certificate has been revoked by the issuing CA.
% In the Must-Staple (MS) configuration before trusting a certificate, the browser is required to obtain proof from the CA that the certificate is still valid and has not been removed yet.
We check for support of CT, DNS CAA records and OCSP in the Must-Staple (MS) configuration.
While all certificates used for the DoH-servers are registered in the CT system, only Google offers DNS CAA records and no server demands OCSP MS. Again, we argue that the introduction of a new secure DNS protocol would be an ideal opportunity to establish and require support for all techniques that can further improve the security of DoH.

% Steve: The following is really weak, so unless we have some really interesting take-home message that is worth repeating, it's just wasting space.

%This section has exposed the variety of implementations of the current DoH server landscape. We believe that this variety constitutes a great opportunity for the research community to further measure and assess this evolving landscape.

%\section{Different transport choices for (secure) DNS}
\section{Transports for secure DNS}

% Moved to previous tex file for placement reasons
%\begin{figure*}[t]
%	\centering
%	\includegraphics[width=\textwidth]{graph/hol-multiplot}
%	\caption{Impact of head-of-line-blocking on resolution times for DNS over different transport protocols. The upper charts depict the baseline and the lower ones the effect of a delay (1000ms for one in 25 queries).}
%	\label{fig:hol-multiplot}
%\end{figure*}

In this section, we investigate the impact of different (secure) transport choices for DNS messages. We compare DNS-over-TLS with DoH using both HTTP/1.1 and HTTP/2.0.\footnote{We do not consider DNScrypt here since it takes an orthogonal approach. Whereas DoT and DoH encapsulate the original DNS UDP wireformat with TLS and HTTP headers respectively, DNScrypt uses a redesigned wireformat combining all these features into a single message.}

% Steve: Given that we measure performance and that it might depend on our setup, a very short description of the setup/hardware/software might be necessary.
We compare the effect of these different choices of transport via a controlled experiment. We set up a local CoreDNS resolver, and use it to resolve 100 domain names via UDP, TLS, HTTP/1.1 and HTTP/2.0.
As we are evaluating the impact of the transport protocol, we instruct our resolver to always return the same IP address independently of the domain name. Using unique domain names for each query rules out any impact of caching while still being able to attribute differences in resolution time to the transport protocol instead of the resolved domain name.
We construct the queried domain names with a random prefix of constant length five  followed by a fixed base domain.
This construction ensures that effects of compression of query names are uniformly distributed across all queries, hence ensuring that differences in compressability of domain names do not impact our results.
To introduce workload variability, we use non-deterministic query arrival time, where query arrival times follow a Poisson distribution with an average arrival rate of 10 queries per second.
Experiments were carried out on a machine running \mbox{CentOS 7} on a 4-core Intel Core i5-2500K CPU (3.30 GHz) and 8GB of RAM.
Python 3.6 and CoreDNS 1.2.2 were used.
Experiments were isolated with Docker containers.
Python's standard packages for sockets, TLS and HTTP/1.1 were used, DNS handling was done with the dnspython package, for HTTP/2.0 support Facebook's doh-proxy package was used.

We carry out two measurement runs. In the first run, we obtain a baseline of the achievable performance by answering queries as fast as possible. We then instruct our resolver to delay one in every 25 queries by 1000ms, to observe whether delays in resolution time affect subsequent answers.

Figure~\ref{fig:hol-multiplot} provides the results in both scenarios for each transport protocol.
Resolution time is the time it takes the application to receive and fully parse a reply, not just the time it takes the network to transmit the message.
The upper row shows baseline performance without delay. The second row shows per query resolution times with the introduced delay.
%From left to right, the columns show results for transport via UDP, TLS, HTTP/1.1 and HTTP/2.0.
The HTTP/1.1 scenario employs HTTP request pipelining, as we are assessing the resilience against slow or delayed queries of the individual transports (so HTTP/1.1 without pipelining would be an unfair comparison).

%Without pipelinig, HTTP/1.1 cannot recover from these, it would thus be unfair to compare against HTTP/1.1 without pipelining, hence we use HTTP/1.1 with pipelining.

For the baseline case without delay, we observe that both UDP and TLS deliver responses to queries in less than one millisecond. These values are expected for a controlled experiment setup running on the localhost.\footnote{The first query is slower than the following ones due to additional packet round-trips by TCP and TLS handshakes during connection setup.}
HTTP/2.0 consistently delivers results in less than ten milliseconds. %, which points to a higher overhead caused by the additional HTTP headers and server state to manage HTTP/2.0.
% Steve: Here I suspect some reviewers may go as far as ask us to break down the overhead since it's a controlled experiment.
Only for HTTP/1.1, the baseline performance fluctuates significantly, which we attribute to issues in the pipelining support. Most major browsers tried to support HTTP/1.1 pipelining, but have ceased to support it due to too many interoperability issues negatively affecting performance~\cite{mozillahttp11,chromehttp11}.

In the bottom row experiments, we observe that DNS via UDP is hardly affected by the delay. We clearly see four outliers for the four delayed queries, without any visible impact on subsequent queries. Indeed, DNS via UDP can utilize different connections via multiple port numbers to effectively multiplex queries and thus make them independent.
%This is possible since UDP is a connection-less protocol that does not require state to be held by the DNS server for the connection.

For TLS as transport protocol, we see that the delayed queries have a knock-on effect on subsequent queries; the serialization of the TLS connection implies that a reply to a subsequent query is only sent after the reply to the delayed query.
Out-of-order delivery of replies via TLS is also possible since every request/reply pair can be identified by their unique ID. The DNS-over-TLS (DoT) RFC states that this feature should be supported, although it is not mandatory. In practice, out-of-order delivery greatly complicates the server implementation compared to standard UDP, as it requires state management on the server side to handle these requests.
From the only three existing DoT servers in the wild (as per Wikipedia at the time of writing this paper), we have verified that only Cloudflare supports out-of-order delivery. Implementing out-of-order delivery via TLS is akin to (re-)implementing the stream multiplexing part of SCTP, QUIC or HTTP/2.0. We believe that this is one of the main reasons why DoT failed to gain significant traction. Surprisingly, we are not aware of any other work explicitly demonstrating the potential performance impact of simple TLS transport on DNS.

For HTTP/1.1, we observe similar knock-on effects as for TLS. In the case of HTTP/1.1, there is no way to circumvent these as the in-order-delivery of requests is demanded by the RFC~\cite{rfc2616}.
It is only when we turn to HTTP/2.0 that we observe a similar insensitivity to delayed queries as with UDP.
Indeed, the DoH RFC states that HTTP/2.0 is the minimum recommended version of HTTP to be used.

In this section we have seen that DNS-over-TLS and DNS-over-HTTPS/1 suffer from head-of-line-blocking.
Only with HTTP/2.0 DoH manages to provide similar results to UDP DNS with respect to head-of-line blocking.
This difference in behavior might (at least in part) explain why it was easier for DNS-over-HTTPS/2.0 to gain traction than for DNS-over-TLS.

\section{Overhead}

\begin{figure}[t]
	\centering
	\includegraphics[width=\columnwidth]{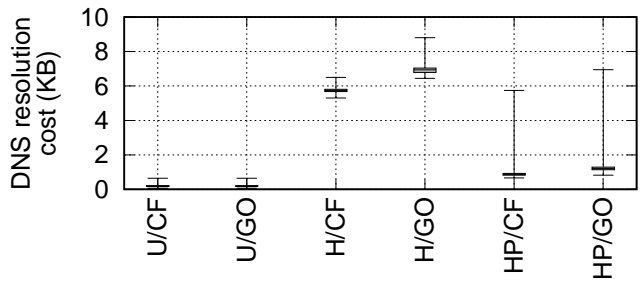}
	\caption{Total bytes per resolution. Domain names were resolved via UDP-DNS (U), DNS-over-HTTPS without persistent connection (H) and with a persistent connection (HP). The DNS servers of Cloudflare (CF) and Google (GO) were used. Whiskers span the full range of values.}
	\label{fig:boxplot-doh-total-bytes-per-resolution}
\end{figure}

%\begin{figure}[t]
%	\begin{center}
%		\subfloat[Bytes per resolution.]{
%			\includegraphics[width=0.2\textwidth]{graph/boxplot-doh-total-bytes-per-resolution}
%			\label{fig:boxplot-doh-total-bytes-per-resolution}
%		}
%		\subfloat[Packets per resolution.]{
%			\includegraphics[width=0.2\textwidth]{graph/boxplot-doh-total-pkts-per-resolution}
%			\label{fig:boxplot-doh-total-pkts-per-resolution}
%		}
%	\end{center}
%	\caption{Total bytes and packets per resolution. Domain names were resolved via UDP-DNS (U), DNS-over-HTTPS without persistent connection (H) and with a persistent connection (HP). The DNS servers of Cloudflare (CF) and Google (GO) were used. Whiskers span the full range of values.}
%	\label{fig:boxplot-bytes-packets-per-resolution}
%\end{figure}

In the previous section, we have seen that DNS-over-HTTPS/2 offers significant advantages over DNS-over-TLS and DNS-over-HTTPS/1.
However, the requirement for HTTP/2 introduces additional layers and thus more headers and overhead. In this section, we compare the overhead incurred by DNS-over-HTTPS/2 and regular UDP-based DNS.

To obtain a set of domain names that is representative of the real-world, we fetch the top 100,000 webpages as per global Alexa ranking and gather all domains that were resolved during these fetches. We instruct the local stub resolver to log all queries. The Alexa list was retrieved on 15 September 2018.
In contrast to browser-generated HTTP Archive (HAR) files, this allows us to obtain the domains that are not part of the actual webpage but are contacted by common web browsers during page load, e.g., OCSP records for secure TLS connection establishment. While fetching these 100,000 webpages, 2,178,235 DNS queries were sent.
As domain names can be embedded in more than one page, these 100,000 page fetches resolved 281,414 unique domain names. Notably, almost 25\% of all DNS queries can be attributed to the fifteen most frequently queried domain names. We then resolve these domain names from a university vantage point via regular UDP-based DNS and DNS-over-HTTPS, using the respective resolvers of both Google and Cloudflare.
%In the case of DoH, we consider both scenarios: without a persistent connection (i.e., single TLS connection per query) and with a persistent connection (i.e., reusing TLS connection to send as many queries as the servers allows).

\begin{figure}[t]
	\centering
	\includegraphics[width=\columnwidth]{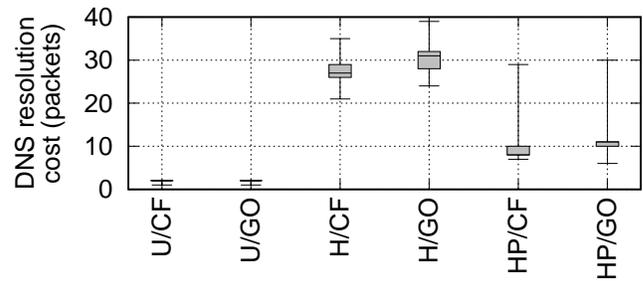}
	\caption{Total packets per resolution. Domain names were resolved via UDP-DNS (U), DNS-over-HTTPS without persistent connection (H) and with a persistent connection (HP). The DNS servers of Cloudflare (CF) and Google (GO) were used. Whiskers span the full range of values.}
	\label{fig:boxplot-doh-total-pkts-per-resolution}
\end{figure}

\begin{figure*}[t]
	\centering
	\includegraphics[width=\textwidth]{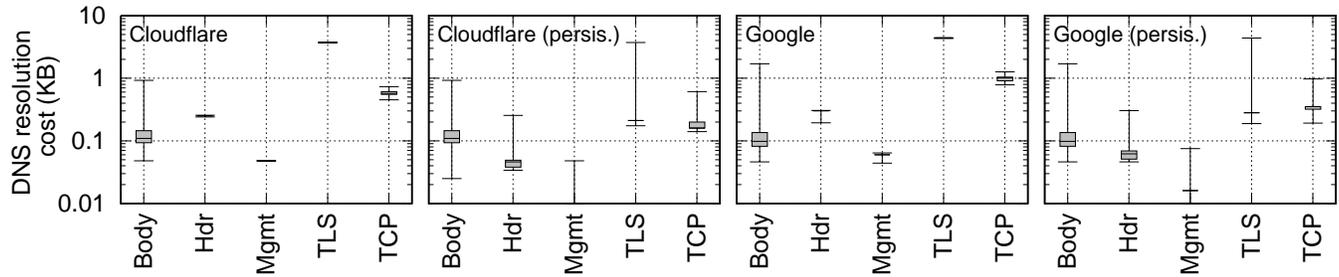}
	\caption{Overheads per DNS resolution for DNS-over-HTTPS/2. First two columns show sizes for (HTTP) bodies and headers exchanged. Mgmt refers to messages being exchanged to maintain the HTTP/2 connection like settings and windows updates. TLS and TCP refer to sizes of the respective layers.}
	\label{fig:boxplot-doh-overheads-multiplot}
\end{figure*}

Figure~\ref{fig:boxplot-doh-total-bytes-per-resolution} shows the distribution of request sizes for all six scenarios.
Figure~\ref{fig:boxplot-doh-total-pkts-per-resolution} depicts the number of packets. % that need to be exchanged to successfully resolve a domain name.
When comparing UDP-based DNS with DoH, we see that the UDP transport systematically leads to fewer bytes and fewer packets exchanged, with the median DNS exchange consuming only 182 bytes bytes and 2 packets. A single DoH resolution in the median case on the other hand requires 5737 bytes and 27 packets to be sent for Cloudflare and 6941 bytes and 31 packets for Google. A single DoH exchange thus consumes more than 30 times as many bytes and roughly 15 times as many packets than in the UDP case. Persistent connections allow to amortize one-off overheads over many requests sent. In this case, the median Cloudflare resolution consumes 864 bytes in 8 packets, the median Google resolution 1203 bytes in 11 packets. While this is significantly smaller compared to the case of a non-persistent connection, DoH resolution still consumes roughly more than four times as many bytes and packets than UDP-based DNS does.

While in the legacy case there is no significant difference, in the DoH case Google's server leads to larger transactions than Cloudflare's.
%In the legacy case, there is no significant difference between Cloudflare and Google. In the DoH case however, the requests are not only systematically larger, but Google's DoH servers also lead to larger transactions than Cloudflare's one.
This is caused by Google needing more bytes to establish and maintain the TLS connection than Cloudflare. The reason is Google's usage of a certificate larger than Cloudflare's: in our specific setup, Cloudflare transmits two certificates worth 1,960 bytes and Google transmits two certificates worth 3,101 bytes. When using a persistent TLS connection, the overheads get amortized over the many requests made.

We now break down the overhead for DoH. As a by-product, we showcase some of the new features of HTTP/2 in comparison to HTTP/1.
Next to header compression using HPACK~\cite{rfc7541}, HTTP/2 also supports a differential transmission mechanism that only transmits the changed headers during the subsequent exchanges. HTTP/2 also defines new message frames to manage the connection.
%Among others, there are frames defined for stream management, updates of the flow control window, ping messages to keep a connection alive and \texttt{goaway} message to initiate connection teardown.
Figure~\ref{fig:boxplot-doh-overheads-multiplot} shows a breakdown of overheads into individual layers and protocols.
%The columns from left to right depict sizes of the HTTP bodies and headers.
%The next columns shows the total size of messages being exchanged to maintain a HTTP/2 connection, e.g., HTTP/2 \texttt{settings}, \texttt{ping} and \texttt{window\_update} messages. The TLS and TCP columns depict the size of the headers for the respective layers.
Across all four cases, the distribution of body sizes is similar, albeit Google tends to send slightly larger bodies in the extreme case.

Every additional layer of complexity adds overhead that is at least the same size as the original DNS payload.
Notably, even the overhead incurred by TLS encryption and TCP headers and additional messages is already of the size of the complete DNS payload.
Regarding the HTTP/2 overhead (headers and mgmt), we see that using a persistent connection leads to less data being exchanged.
For the headers, this is caused by HTTP/2's differential headers feature, which in sequential requests and replies only transmits those headers that have changed. %are different from the previous reply or request respectively.
The management messages are required to manage the HTTP/2 connection and multiplexing of different streams.
They do not need to be sent for every single client-server-interaction.
Therefore, when using a persistent and thus re-usable connection, the amount of management bytes sent per request-response-cycle is smaller in comparison to non-persistent connections.
For the overhead incurred by TLS, for the non-persistent connections, the overhead is dominated by the server certificate as discussed above.
In the case of persistent connections, the upper whiskers in Figure~\ref{fig:boxplot-doh-overheads-multiplot} are caused by the (at least once) necessary certificate exchange.
The median values however are significantly lower as an established connection is re-used many times.
This variability in the TLS overhead also causes different overheads at the TCP and outer layers, as the higher number of bytes transmitted for the TLS layer also leads to more packets.

In summary, many of the one-time overheads required for TCP, TLS and HTTP connection setup and management can be amortized if a persistent connection is used. However, even in this case, the median overhead caused by the TLS and TCP layer are each already of the size of the actual DNS message. For DNS resolution over HTTP, this effect is pronounced because of the comparably small size of the DNS message.
When considering transmitting web pages via HTTPS, this effect will be less pronounced in comparison to DNS messages, given the larger size of websites.

\section{DoH Performance}

\begin{figure*}[t]
        \centering
        \includegraphics[width=\textwidth]{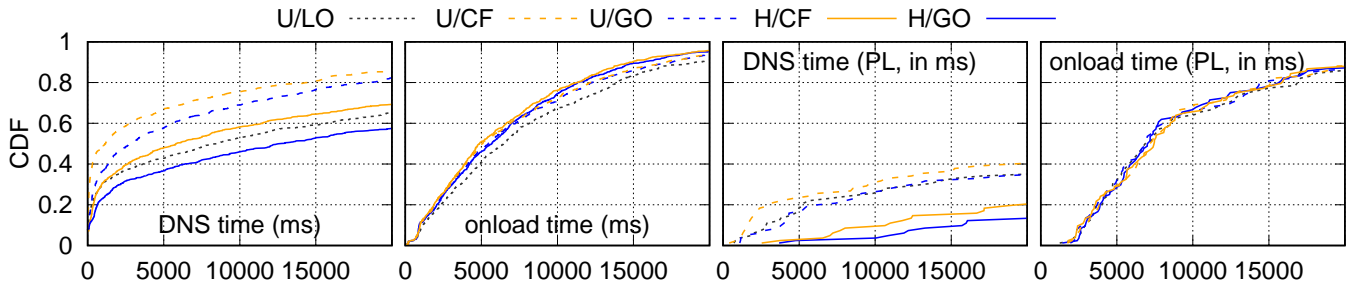}
        \caption{CDF of DNS resolution and page load times (time of onload event): U/ indicates legacy resolver, H/ indicates resolution via DoH, /LO indicates local resolver, /GO indicates Google and /CF indicates Cloudflare.}
        \label{fig:cdf-pageload-multiplot}
\end{figure*}

In the previous sections we have quantified the potential impact of head-of-line-blocking as well as the additional overheads of DoH.
In this section, we assess whether DoH impacts performance, more specifically we look at a web browsing scenario and investigate how a change to DoH affects page loading times.

We use the Firefox web browser to measure webpage load times for the 1,000 highest ranked webpages in the global Alexa ranking.
The Alexa list was retrieved on 18 April 2019.
We choose Firefox because as of the time of writing this paper it was the only browser with documented support for DoH.
We use Firefox 66.0.3 for the experiments.
We rely on the Browsertime framework from the sitespeed.io project\footnote{\url{https://www.sitespeed.io}} to instruct Firefox for the measurements and collect HAR files with the performance statistics.

We measure performance using the locally configured resolver, and also using the public resolvers from Google and Cloudflare over legacy DNS as well as DoH.
This way, the performance obtained with the local resolver provides a baseline, allowing us to assess how a change to a cloud-provided DNS service affects performance.
For the cloud provided DNS services, we also assess the performance difference between using the traditional UDP-based DNS protocol and DNS-over-HTTPS.
In this setup, each website was loaded three times with the browser cache purged before each measurement iteration. This was done from a university-local server.

The left plot in Figure~\ref{fig:cdf-pageload-multiplot} shows the CDFs of the cumulative DNS resolution times per webpage in milliseconds. By cumulative DNS resolution times, we mean the time it would take to perform all DNS queries serially, whereas in reality they can be parallelised.
We crop the CDF plot at 20,000ms, since the results have a very long tail.
%This capped tail also explains why the CDFs do not cover the full range of values up to one.
%We have further decided to plot all CDFs in the same axis layout to allow for easier comparison of them, which is why some plots, especially the third CDFs, only show comparibely small values.

We first observe that the cloud-based name resolution via UDP leads to faster resolution times than using the local resolver.
%Using the Cloud resolvers, at most 5,000 ms are required for name resolution for more than half of the pages, whereas for the local resolver this is the case for less than 45\% of the pages.
% Steve: The previous sentence is weird/bad, it's making a comparison where there is little difference, while the tone suggests it's trying to imply there is something in here, which will lead the reviewer to say: MEH, BORING! What's the point of mentioning this at all?
From the cloud-based ones, Cloudflare leads to faster resolution times than Google.
When comparing DoH-services from Cloudflare and Google, we observe that using DoH leads to longer DNS resolution times than when using the traditional DNS resolution.
This is to be expected from the added overhead for encryption and transport.
Also, we observe that the DoH resolution provides comparable resolution times to the local resolver, with again Cloudflare slightly faster than Google.

Even though these results show that changing to DNS resolution via DoH leads to longer DNS resolution times, this does not necessarily translate into longer page load times.
The second plot in Figure~\ref{fig:cdf-pageload-multiplot} shows CDFs of the complete page load time, measured as the time when the onload event was triggered.
The onload event is triggered when the whole page including all dependent resources like stylesheets and images has been loaded~\cite{mdn_load_event}.
Note that overall page load times are faster than DNS resolution times as the browser sends requests in parallel, whereas DNS plots shows cumulative DNS resolution times without parallelism.
% Steve: previous sentence is not precise technically as what is really being measured (what is begin and end of the load time?)
The figure shows that page load times are comparable for all resolution approaches.
As for the previous DNS resolution times, using a cloud-based DNS service offers slightly faster page load times.
There is however little difference between page load time via legacy DNS or DNS-over-HTTPS: both resolution mechanisms achieve similar page load times.

Note that we also attempted to run the same experiments from PlanetLab. Unfortunately, at the time of writing this paper, only 39 nodes were able to run these experiments, as most of them were unreachable, and among those that were reachable, many were running an OS that was too old to support a recent enough version of Firefox that supports DoH. The limited results (plots on the right in Figure~\ref{fig:cdf-pageload-multiplot}) we obtained from PlanetLab however are consistent with those we have obtained locally: DNS resolution via DoH takes longer, but page load times overall change only little when changing the resolution method.

Overall, the results of this section show that a switch to DNS-over-HTTPS does not seem to incur significantly longer page loading times.
This means it is feasible to benefit from the better privacy guarantees of DoH without sacrificing user-perceived page loading times.

\section{Related Work}
DNS-over-HTTPS still is a relatively new protocol.
%, recently proposed and still undergoing standardization.
To the best of our knowledge, this paper is the first to look into the differences between DNS-over-HTTPS, DNS-over-TLS and UDP-based DNS.
Mozilla has published a blog post~\cite{mozillablog} briefly describing their experience with a DoH trial in Firefox. This blog post however focuses more on reporting experiences of using a third-party resolver than on implications that stem directly from using DoH, especially the transport aspect.
In a blog post~\cite{apnicdoh}, Geoff Huston also asks for the advantage of DoH over DoT. This post discusses application features like HTTP push and namespaces, but does not discuss insensibility against slow queries as we do.
%While DoH implies an improvement in terms of DNS security, the distinct features of DNS traffic still allow to identify the presence of websites in traffic traces~\cite{siby2018dns}.
% Steve: So what? Give the punch if there is one instead of leaving it at the level of the unclear implication. WHAT DOES THIS IMPLY?

Since the inception of DNS, the Internet has evolved and changed, exposing the DNS protocol to new threats and challenges. The unencrypted transport of DNS leads to security and censorship issues~\cite{bortzmeyer2015dns,levis2012collateral}, whereas using UDP makes DNS usable for distributed denial-of-service attacks~\cite{anagnostopoulos2013dns}.
Other works have proposed protocol changes to use persistent connections and encryption~\cite{zhu2015connection}.
These works list and discuss issues with the traditional UDP-based transport for DNS, of which most can be addressed by using DNS-over-HTTPS instead. In that sense, they provide good arguments to change to DoH, but do not discuss details of DoH directly.

Content delivery networks often use DNS to perform their traffic redirection. It is an active research area, with works aiming at better understanding these redirection strategies~\cite{bottger2018open,calder2013mapping,otto2012content}. Other works study DNS resolver behavior in the wild with respect to latency and traffic redirection~\cite{ager2010comparing}, look at the impact of DNS on overall application delays in the Internet~\cite{bozkurt2017internet,sundaresan2013web} or look at DNS infrastructure provisioning at the client side~\cite{schomp2013measuring}. While all these works also target DNS, they have a stronger focus on the actual applications of DNS than the protocol itself.

\section{Conclusion}

DNS is one of the most important protocols for many networked applications today and was originally designed as an unencrypted protocol. Growing concerns about user privacy have led to propose more secure approaches.
%To achieve this, one approach is to send DNS requests over secure transport protocols, such as DNS-over-TLS (DoT) or DNS-over-HTTPS (DoH).
In this paper, we have surveyed the current DoH landscape. We have exposed the diversity in the supported content types, in the support for DNS-over-TLS, and in the supported TLS versions. We have seen, that while most DoH servers support a good set of security parameters, many of them still do support deprecated legacy settings. We have then studied the behavior of DoT and DoH against delayed queries, showing that HTTP/2 offers advantages over HTTP/1 and DNS-over-TLS. In the process, we have exposed the likely reason why DoT has not gained traction compared to DoH, despite having had a head start of a few years before DoH. We have then quantified the overheads incurred by the HTTP and TLS layers of HTTP/2. Finally, we have measured how DoH impacts page load times. This has shown that it is possible to obtain the additional security of DoH with only marginal performance penalties.

\begin{acks}
We thank our shepherd Taejoong Chung and the anonymous reviewers for their reviews and constructive feedback.

This research is supported by the UK's \grantsponsor{epsrc}{Engineering and Physical Sciences Research Council (EPSRC)}{https://epsrc.ukri.org/} under the \grantnum{epsrc}{EARL: sdn EnAbled MeasuRement for alL project (Project Reference EP/P025374/1)}.
\end{acks}

\bibliographystyle{ACM-Reference-Format}
\bibliography{paper}

\end{document}